\documentclass[final,5p,times,twocolumn,authoryear]{elsarticle}

\usepackage{amssymb}
\usepackage{lipsum}
\usepackage{amsmath}
\usepackage{threeparttable}

\usepackage{natbib}
\usepackage{hyperref}

\journal{Journal of High Energy Astrophysics}

\begin{document}

\begin{frontmatter}



\title{Constraining the initial Lorentz factor of gamma-ray bursts under different circumburst mediums}


\author[1]{Sheng-Jin Sun}
\author[1]{Shuang-Xi Yi*}
\author[2]{Yuan-Chuan Zou**}
\author[1]{Yu-Peng Yang}
\author[3]{Ying Qin}
\author[4]{Qing-Wen Tang}
\author[5]{Fa-Yin Wang}

\address[1]{School of Physics and Physical Engineering, Qufu Normal University, Qufu, 273165, Shandong, China; yisx2015@qfnu.edu.cn}
\address[2]{School of Physics, Huazhong University of Science and Technology, Wuhan, 430074, Hubei, China; zouyc@hust.edu.cn}
\address[3]{Department of Physics, Anhui Normal University, Wuhu, 241002, Anhui, China}
\address[4]{Department of Physics, Nanchang University, Nanchang, 330031, Jiangxi, China}
\address[5]{School of Astronomy and Space Science, Nanjing University, Nanjing 210023, China}

\begin{abstract}
The initial Lorentz factor ($\Gamma_{\text{0}}$) plays a crucial role in uncovering the physical characteristics of gamma-ray bursts (GRBs). Previous studies have indicated that the ambient medium density index $k$ for GRBs falls in the range of 0 - 2, rather than exactly equal to 0 (homogeneous interstellar ambient) or 2 (typical stellar wind). In this work, we aim to constrain the $\Gamma_0$ of GRBs considering their distinct circumburst medium. We select a total of 33 GRBs for our analysis, comprising 7 X-ray GRBs and 26 optical GRBs. Subsequently, by utilizing the deceleration time of fireball $t_{\rm p}$, we derive the $\Gamma_0$ for the 33 GRBs assuming the radiation efficiency of $\eta =$ 0.2. The inferred initial Lorentz factor was found to be from 50 to 500, consistent with previous studies. We then investigate the correlation between the $\Gamma_0$ and the isotropic energy $E_{\rm \gamma,iso}$ (as well as the mean isotropic luminosity $L_{\rm \gamma,iso}$), finding very tight correlations between them, i.e., $\Gamma_0$ $\propto$ $E^{0.24}_{\rm \gamma,iso,52}$ ($\Gamma_0$ $\propto$ $L^{0.20}_{\rm \gamma,iso.49}$) with $\eta$=0.2. Additionally, we verify the correlation among $\Gamma_0$, the isotropic energy $E_{\rm \gamma,iso}$ (or $L_{\rm \gamma,iso}$) and the peak energy $E_{\rm{p,z}}$, i.e., $E_{\rm \gamma,iso,52}$ $\propto$ $\Gamma^{1.36}_0$$E^{0.82}_{\rm{p,z}}$ ($L_{\rm \gamma,iso,49}$ $\propto$ $\Gamma^{1.05}_0$$E^{0.66}_{\rm{p,z}}$) under the same radiation efficiency ($\eta$=0.2).
\end{abstract}




\end{frontmatter}




\section{Introduction}
\label{Sec:A: Introduction}

Gamma-ray bursts (GRBs) are the most powerful and violent explosions in the universe. The relativistic fireball model is one of the few models that has been put forth to explain the origin of GRBs (i.e., \citealp{2002ARA&A..40..137M}; \citealp{2004RvMP...76.1143P}; \citealp{2004IJMPA..19.2385Z}). In the standard fireball model scenario, two types of shocks are generated when the fireball shell material interacts with the surrounding interstellar medium: the forward shock (FS) through the interstellar medium and the reverse shock (RS) sweeping through the shells of the fireball (\citealp{1995ApJ...455L.143S}; \citealp{1997ApJ...476..232M}; \citealp{1998ApJ...497L..17S}; \citealp{2000ApJ...545..807K}; \citealp{2013ApJ...776..120Y}; \citealp{2014ApJ...792L..21Y}). The observed afterglow is generated through synchrotron radiation, arising from the interaction between the fireball and the surrounding medium. The duration of a GRB is commonly denoted by $T_{90}$, where $T_{90}$ $ > $ 2 s corresponds to long GRBs, and $T_{90}$ $ < $ 2 s refers to short GRBs (\citealp{2013ApJ...763...15Q}; \citealp{2020ApJ...893...77W}). Generally, short GRBs originate from the merger of binary compact objects (\citealp{2002ApJ...571..779P}), while long GRBs result from the collapse of massive stars (\citealp{1993ApJ...405..273W}; \citealp{1999ApJ...524..262M}). Recent observations challenge the traditional classification of GRBs. For example, GRB 200826A, classified as a short GRB based on its duration, exhibits observational evidence consistent with originating from the collapse of a massive star, thereby challenging the notion that all short GRBs are exclusively generated by binary compact object mergers (\citealp{2022ApJ...932....1R}).

The two types of GRBs are considered to have distinct circumburst medium environments, which have been constrained in previous studies by employing GRB afterglow light curves. In general, the density of their circumburst medium for long GRBs can be expressed as $n$ = $AR^{-k}$ ($R$ is the radius and the $k$ is the medium density distribution index). Here $k$ = 0 and $k$ = 2 represent a homogeneous interstellar medium (ISM) and a typical stellar wind environment, respectively. Recently, \cite{2022ApJ...925...54T} collected a sample of 18 X-ray afterglows that clearly show notable FS characteristics. They further inferred the value of $k$ to be in the range of 0.2 - 1.8, assuming the external FS model. Their results indicate that the ambient medium of those GRBs is mostly between ISM and the stellar wind, which is consistent with earlier findings \citep{2013ApJ...774...13L,2013ApJ...776..120Y,2020ApJ...895...94Y}.

It's worth mentioning that previous studies most have been focused on the two cases: $k$ = 0 (ISM) (\citealp{1998ApJ...497L..17S}; \citealp{2012ApJ...751...57D}; \citealp{2013NewAR..57..141G}) or $k$ = 2 (typical stellar wind environment) (\citealp{1998MNRAS.298...87D}; \citealp{2000ApJ...536..195C}; \citealp{2003MNRAS.342.1131W}; \citealp{2005ApJ...619..968W}; \citealp{2005MNRAS.363...93Z}; \citealp{2012ApJ...751...57D}). In the canonical fireball model, the evolution of the fireball proceeds through three distinct phases: acceleration, coasting, and deceleration. During the acceleration phase, the fireball is dominated by radiation, with its acceleration primarily driven by radiative pressure. The Lorentz factor increases with the expansion of the fireball radius in this phase. Upon completion of the acceleration phase and transition into the coasting phase, the fireball is no longer significantly influenced by radiative pressure and instead moves with a relatively constant Lorentz factor. This constant Lorentz factor, which is the Lorentz factor at the commencement of the coasting phase, is designated as the initial Lorentz factor ($\Gamma_0$). Throughout the coasting phase, the fireball maintains this approximately constant Lorentz factor ($\Gamma_0$) until it begins to interact with the circumambient medium, thereby entering the deceleration phase (\citealp{2018A&A...609A.112G}; \citealp{2019ApJ...883...97Z}). The initial Lorentz factor plays a crucial role in understanding the physical characteristics of GRBs. However, obtaining its accurate value is facing challenges.

To date, three methods have been usually proposed to estimate the initial Lorentz factor $\Gamma_0$. The first method, known as the ``compactness problem'' argument, uses the high-energy cutoff in the spectrum as an indicator of pair-production opacity to constrain the value of $ \Gamma_0$ \citep{1997ApJ...491..663B,2001ApJ...555..540L,2008MNRAS.384L..11G,2009Sci...323.1688A,2024A&A...685A.166R}. However, the compactness argument primarily provides a lower limit on $ \Gamma_0$, and this method relies on the assumption of the internal shock radius $R_{\gamma}$ = $\Gamma^2$$c$$\delta$$t$. Here, $\Gamma$ represents the Lorentz factor, $c$ is the speed of light, and \textbf{$\delta$$t$} is the variability time scale. The minimum variability time scale \textbf{$\delta$$t$} is indeed subject to large uncertainty due to the chaotic nature of GRB light curves. The second method involves utilizing the blackbody component in the GRB spectrum and is applicable exclusively to a subset of GRBs that exhibit such components within their detected spectrum \citep{2007ApJ...664L...1P,2014ApJ...795..155P,2015ApJ...800L..23Z}. The limitation of this approach are that it requires modeling the prompt emission of the fireball and (\citep{2015ApJ...801..103G}), more importantly, it requires an identification of the thermal composition. However, some recent works in literature suggest that the claimed thermal components in GRB spectra may not real, and can be interpreted instead within a marginally fast cooling synchrotron model (\citealp{2017ApJ...846..137O};\citealp{2018A&A...613A..16R};\citealp{2021A&A...652A.123T};\citealp{2023ApJ...959...91P};\citealp{2025A&A...693A.320D}). The third method involves analyzing the smooth onset bump in early afterglow light curves, i.e., using the peak time$\ t_{p}$ of the bump as the deceleration time of fireball to infer $ \Gamma_0$ \citep{1999ApJ...520..641S,2006ApJ...642..354Z,2010ApJ...725.2209L,2018A&A...609A.112G}. Considering the limitations and drawbacks of the first two methods, the third method has become the most widely accepted approach.

With the success of the constraint of the initial Lorentz factor, several interesting correlations with other prompt emission properties (the isotropic energy $ E_{\rm \gamma,iso}$, the isotropic equivalent peak luminosity $ L_{iso}$, the mean isotropic luminosity $L_{\rm \gamma,iso}$, the peak energy $ E_{\rm{p,z}}$ and so on) can thus be revealed (\citealp{2005ApJ...633..611L}; \citealp{2008MNRAS.388.1284R}; \citealp{2010ApJ...725.2209L}; \citealp{2012ApJ...751...49L}; \citealp{2012MNRAS.420..483G}; \citealp{2018A&A...609A.112G}). By constraining the initial Lorentz factor of the GRB with onset bump feature, \cite{2010ApJ...725.2209L} found a tight correlation of $\Gamma_0$$\simeq$195$E^{0.27}_{\rm \gamma,iso,52}$. Based on this, \cite{2012ApJ...751...49L} investigated the an even tighter correlation between the initial Lorentz factor $\Gamma_0$ and the mean isotropic luminosity $L_{\rm \gamma,iso}$. Additionally, several multi-variable correlations have been proposed, such as, $ \Gamma_0$ - $ E_{\rm \gamma,iso}$ - $ E_{\rm{p,z}}$ and $ \Gamma_0$ - $ L_{\rm \gamma,iso}$ - $ E_{\rm{p,z}}$. These relationships offer valuable insights into the underlying physical mechanisms of GRBs.

Therefore, it is necessary to investigate the initial Lorentz factor $ \Gamma_0$ for different circumburst mediums, and further explore the potential correlations of $ \Gamma_0$ with other parameters. In this paper, we endeavor to infer the $ \Gamma_0$ for GRBs under various circumburst mediums by the peak time $t_{\rm{p}}$, and investigative the correlations of the $ \Gamma_0$ with $E_{\rm \gamma,iso}$, $L_{\rm \gamma,iso}$ and $E_{\rm{p,z}}$. The structure of this article is as follows. We first introduce the sample selection for GRBs with the onset bump feature of their afterglow light curves and relevant data collection in Section \ref{Sec: sample}. In Section \ref{Sec: 3}, we then compute the initial Lorentz factors for the 34 selected GRBs. Using the obtained $\Gamma_0$, in Section \ref{Sec: 4}, we present the correlations $\Gamma_0$ - $E_{\rm \gamma,iso}$ (or $L_{\rm \gamma,iso}$) and $E_{\rm \gamma,iso}$ (or $L_{\rm \gamma,iso}$) - $\Gamma_0$ - $E_{\rm{p,z}}$. The conclusions and discussions are presented in Section \ref{Sec: conclusions}.

\section{Sample selection and data collection}
\label{Sec: sample}
A smooth bump feature appears in the optical afterglow light curves of some GRBs and similar features are also found in X-ray afterglows. Previous studies have applied RS or FS models to GRBs with clear bump features. For instance, with a sample of 19 long GRBs with smooth onset peak features, \cite{2013ApJ...776..120Y} found the circumburst medium index $k$ of these GRBs is in the range of 0.4 - 1.4. Similar results have also been obtained for optical afterglow samples with RS model \citep{2020ApJ...895...94Y} and X-ray afterglow samples with FS model \citep{2022ApJ...925...54T}.
These findings indicate that the ambient medium of GRBs is neither a homogeneous interstellar medium nor a typical stellar wind environment.

In this work, we use the medium density distribution index $k$ obtained from earlier studies to further estimate the initial Lorentz factor $\Gamma_0$ under different circumburst mediums. We select 33 GRBs with redshift ($z$) from the three previous studies as our target sample. In total, there are 26 Optical-select samples from \cite{2013ApJ...776..120Y} and \cite{2020ApJ...895...94Y}, and 7 X-ray GRBs from \cite{2022ApJ...925...54T}. By fitting the onset peak appearing in the light curve with the FS model and RS model, they obtained the values of the medium density distribution index $k$ and the electron spectral index $p$ for 33 GRBs, and the results are shown in Table \ref{tab:addlabel}. In Table \ref{tab:addlabel}, we also present the redshifts ($z$), the rise time indices ($\alpha_{1}$), the decay time indices ($\alpha_{2}$), the spectral indices ($\beta$), and the spectral regimes of each sample, as compiled from the literature.

As the most luminous GRB ever detected, GRB 221009A has been extensively studied since its discovery. Here, we consider the possibility of using it as a sample for the present work. \cite{2023Sci...380.1390L} recently released the high energy afterglow lightcurve of GRB 221009A, which exhibits a distinct onset bump feature. In addition, they also obtained the rising index $\alpha_{1}$ = 1.82$\pm$0.20 and the decay index $\alpha_{2}$ = 1.12$\pm$0.01, respectively. However, analyses by the LHAASO Collaboration suggest that this high-energy afterglow is likely not pure synchrotron radiation but instead involves complex non-synchrotron radiation mechanisms, such as external Compton scattering or proton synchrotron emission (\citealp{2024ApJ...973L..44F}; \citealp{2024ApJ...972L..25Z}; \citealp{2025JHEAp..45..392Z}; \cite{2025A&A...693A.290Z}).
Furthermore, the analysis conducted by the LHAASO Collaboration was carried out under the premise that GRB 221009A resides within the thin shell regime. \cite{2013ApJ...776..120Y} presented a method for determining whether a GRB is in the thick shell regime or thin shell regime by utilizing the
parameter $\xi \equiv\left(\frac{l}{\Delta}\right)^{\frac{1}{2}} \Gamma_0^{-\frac{4-k}{3-k}}$, where $ \Delta \sim c T_{90,z}$ is the width
of the shell, and $l = ((3 - k)E_0 / (4 \pi A m_p c^2))^{1/(3 - k)}$ is the Sedov length.
Specifically, a GRB is classified as being in the thick shell regime when $\xi$ is less than 1,
whereas it is categorized as being in the thin shell regime when $\xi$ exceeds 1. We found that the $\xi$ value of GRB 221009A is 0.27,
indicating that GRB 221009A should be in the thick shell region. Consequently, GRB 221009A should be not expected to conform to the conventional
afterglow model, leading us to decide against incorporating it as a sample for the present study.

\begin{figure*}
	\centering
	\includegraphics[width=0.4\textwidth, angle=0]{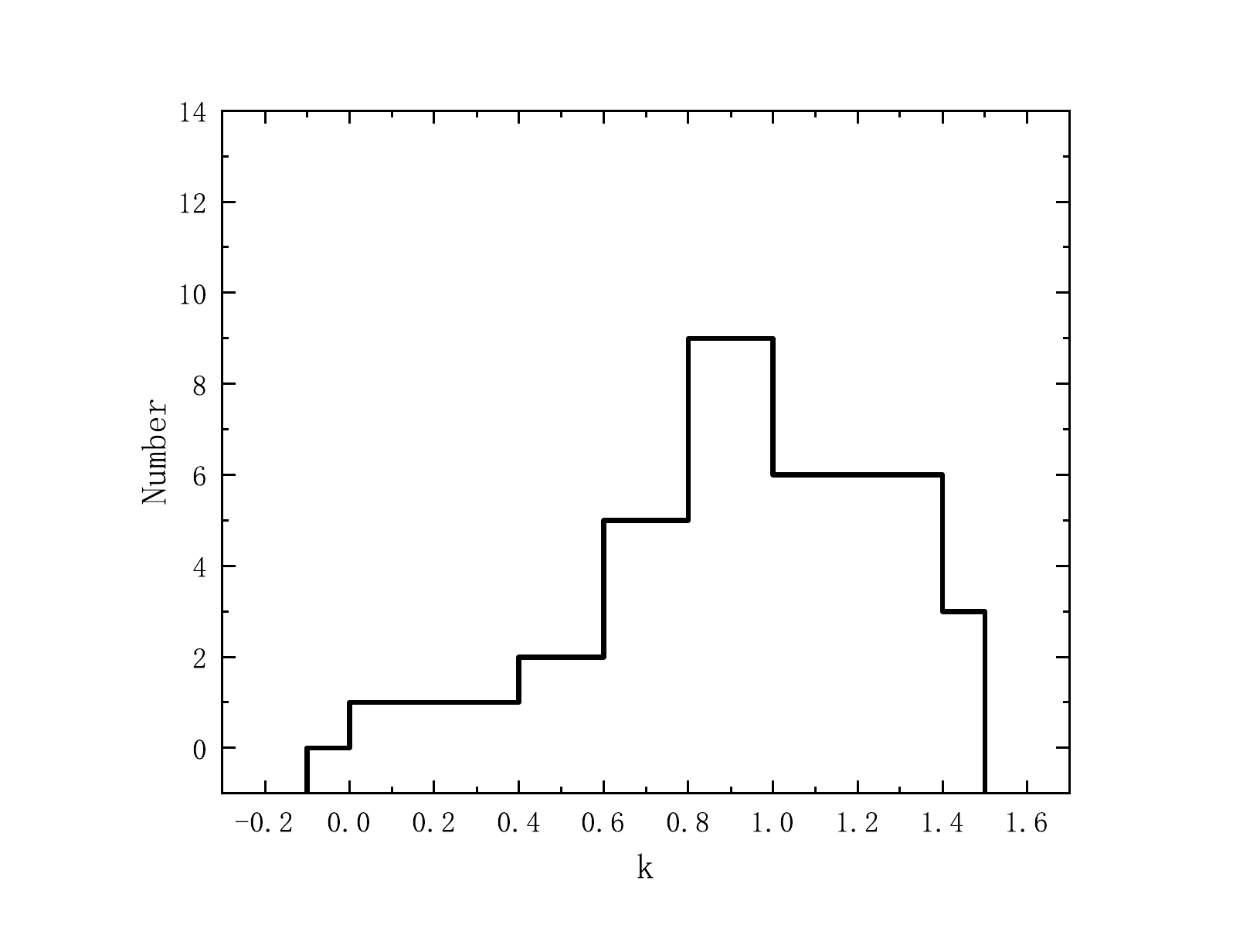}	\includegraphics[width=0.4\textwidth, angle=0]{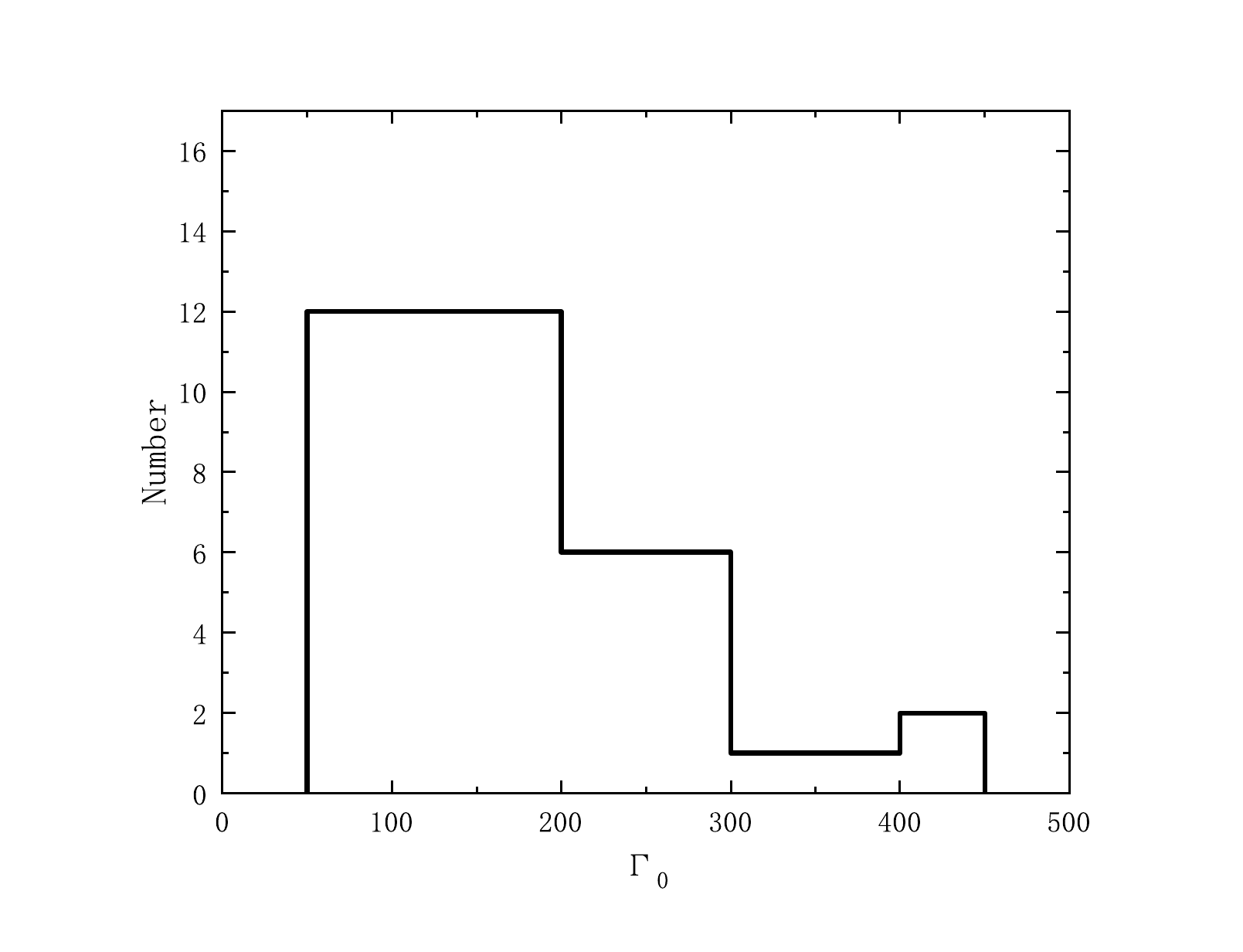}
	\caption{Distributions of the index $k$(left) and $ \Gamma_0$(right) for our selected sample.}
	\label{figure1}
\end{figure*}

\section{Initial Lorentz factor constraints}
\label{Sec: 3}
For the constraints on the initial Lorentz factor, two cases are generally considered: the typical stellar wind environment and the homogeneous interstellar medium \cite[e.g.,][]{2010ApJ...725.2209L,2013ApJ...774...13L,2012MNRAS.420..483G,2018A&A...609A.112G}. \cite{2010ApJ...725.2209L} derived the values of $\Gamma_0\, \sim $ a few hundred for some GRBs in a homogeneous interstellar medium. Previous studies showed that for the density profile $n$ = $AR^{-k}$, the index $k$ is neither 0 (ISM) nor 2 (interstellar wind), but in a range of 0 - 2. With the samples given in Table \ref{tab:addlabel}, we can estimate $\Gamma_0$ for different circumburst medium.

\cite{2013MNRAS.433.2107N} introduced a new framework for modeling the time-dependent behavior of GRB fireball during afterglow phases, applicable to adiabatic systems as well as fully and partially radiative scenarios. Their approach enables the calculation of bolometric afterglow light curves and establishes a revised analytical expression for determining the initial Lorentz factor. The derivation relies on identifying the intersection between luminosity solutions corresponding to the coasting phase and the deceleration phase. This may be a more realistic description of the dynamics for GRB fireball, and therefore we adopt this mothed to constrain the initial Lorentz factor. Given that most samples are in the thin shell regime (duration time $T_{90}$$<$ peak time $t_p$), we refer to \citealp{2018A&A...609A.112G} and use the formula of \citealp{2013MNRAS.433.2107N} to constrain the initial Lorentz factor. The peak time $t_{\rm p}$ in the early afterglow light curve with onset feature can be used for estimating $\Gamma_0$ \citep[e.g.,][]{2013MNRAS.433.2107N}, and the corresponding equation can be written as
\begin{equation}
\Gamma_{0}^{(N 13)}=\left[\frac{(17-4 k)(9-2 k) 3^{2-k}}{2^{10-2 k} \pi(4-k)}\left(\frac{E_{0}}{A m_{\mathrm{p}} c^{5-k}}\right)\right]^{\frac{1}{8-2 k}} t_{\mathrm{p}, \mathrm{z}}^{-\frac{3-k}{8-2 k}},
\label{eq:1}
\end{equation}
where $m_{\rm p}$ is the proton mass, $c$ is the speed of light, and $E_0$ is the kinetic energy of the blast wave. In the scenario of the ambient medium surrounding the burst with a density $n$, a fireball shell with an initial Lorentz factor ($\Gamma_0$) expands outward. $n$ can be expressed in terms of a general power law distribution $n$ = $A$$R^{-k}$ = $n_0$$(R/R_0)^{-k}$. In this distribution, $R_0$ is the value of the radius at $n_0$. Therefore, $A$ in equation \ref{eq:1} can be expressed as

\begin{equation}
A = n_{0}R_{0}^{k}
\label{eq:2}
\end{equation}

By using equation \ref{eq:2} and assuming $n_0 = 1 \,cm^{-3} $(\citealp{2023ApJ...942...81C}), $\Gamma_0$ can be rewritten as

\begin{equation}
\Gamma_0=\left[\frac{(17-4 k)(9-2 k) 3^{2-k}}{2^{10-2 k} \pi(4-k)}\left(\frac{E_0}{R_0^k m_{\mathrm{p}} c^{5-k}}\right)\right]^{\frac{1}{8-2 k}} t_{\mathrm{p}, \mathrm{z}}^{-\frac{3-k}{8-2 k}}
\label{Eq: 2}
\end{equation}

Applying the reverse-forward shocks to 16 optical samples with redshift, \cite{2013ApJ...776..120Y} fitted their afterglow light curves with a smooth bump feature for the standard afterglow model. Additionally, \cite{2020ApJ...895...94Y} adopted the standard afterglow model to 11 optical samples, and derived the value of $R_0$ for each GRB. Here we adopt the fitting results of $R_0$ for our optical sample. As for the X-ray samples, in previous work we only used the rising and decaying slopes of their light curves to constrain the values of $k$ and $p$, without obtaining the
values of $R_0$ (\citealp{2022ApJ...925...54T}). Here we adopt the typical value of $R_0$ as $10^{18}\rm cm$ after consulting various
studies (\citealp{2000ApJ...536..195C,2016A&A...589A..37V,2023ApJ...942...81C}) as shown in Table \ref{tab:addlabe2}.

In both samples, we adopt the radiation efficienciy $\eta$ = 0.2 following the convention established in previous studies \citep{1999ApJ...520..641S,2007A&A...469L..13M,2010ApJ...725.2209L,2013ApJ...774...13L}, where $\eta$ = $E_{\gamma,iso}$/$E_{total}$, and $E_{total}$ = $E_{\gamma,iso}$ + $E_{0}$, $E_{0}$ is the initial isotropically kinetic energy of fireball shell. Here we chose to constrain $E_{0}$ using the radiative efficiency rather than the model of \cite{2013ApJ...776..120Y}. It is mainly due to the model of \cite{2013ApJ...776..120Y} heavy reliance on many uncertain shock microphysics parameters during the light curve fitting process. In contrast, using radiative efficiency to constrain kinetic energy provides a more straightforward and simplified way, as it depends on only one parameter ($\eta$). Hence, using the data in Table \ref{tab:addlabe2}, we further calculate $\Gamma_0$ for 33 GRBs, and the results are presented in Table \ref{tab:addlabe2}. For the error analysis, we used the error propagation formula $\Delta y$ = $\mid \frac{\partial y}{\partial x_1} \mid$$\Delta x_1$ + $\mid \frac{\partial y}{\partial x_2} \mid$$\Delta x_2$ + $\mid \frac{\partial y}{\partial x_3} \mid$$\Delta x_3$ and set $x_1$ = $k$, $x_2$ = $E_{\rm \gamma ,iso}$ and $x_3$ = $t_{p,z}$, which can be used to determine the error in $\Gamma_0$. We show the distribution of $\Gamma_0$ on the right side of Figure \ref{figure1}. The values of $\Gamma_0$ are noted to be within the range of 50-500, which is considered to be within a reasonable range.

It is worth noting that the values of the initial Lorentz factor $\Gamma_0$ constrained by the peak time in this work show some discrepancy with those obtained from afterglow model fitting in previous studies \cite[e.g.,][]{2013ApJ...776..120Y,2020ApJ...895...94Y}. A comparison between the two sets of values reveals differences within a factor of 2 - 4. However, all derived values fall within a reasonable range and are acceptable.

\begin{table*}[htbp]
  \small
  \begin{center}
  \caption{Parameters of our sample.}{
    \begin{threeparttable}
    \begin{tabular*}{0.95\linewidth}{c|c|c|c|c|c|c|c|c}
  \hline
  GRB(band) & $z$ &$\alpha_{1}$   & $\alpha_{2}$   &  k            &  p            & $\beta$ & Emission regime            & References \\
  \hline
  990123(o) & 1.60&$3.31\pm0.25$  & $-3.63\pm0.93$ & $0.71\pm0.12$ & $2.20\pm0.08$ & $0.60\pm0.04$ & $\nu_m^f<\nu<\nu_c^f$        & (1) \\
  041219A(o)& 0.31&$3.50\pm0.21$  & $-1.80\pm0.12$ & $0.80\pm0.10$ & $2.40\pm0.10$ & $0.70\pm0.05$ & $\nu_m^f<\nu<\nu_c^f$        & (1) \\
  050730(o) & 3.97&$1.36\pm0.18$  & $-1.02\pm0.17$ & $0.92\pm0.11$ & $2.16\pm0.23$ & $0.56\pm0.06$ & $\nu_m^f<\nu<\nu_c^f$        & (2) \\
  060605(o) & 3.77&$0.90\pm0.09$  & $-1.17\pm0.05$ & $1.04\pm0.09$ & $2.23\pm0.07$ & $1.04\pm0.05$ & $\nu>max\{\nu_c^f,\nu_m^f\}$ & (2) \\
  060607A(o)& 3.08&$3.21\pm0.11$  & $-2.73\pm0.29$ & $0.69\pm0.08$ & $2.12\pm0.08$ & $0.56\pm0.04$ & $\nu_m^f<\nu<\nu_c^f$        & (1) \\
  060904B(o)& 0.70&$1.10\pm0.16$  & $-0.85\pm0.08$ & $0.95\pm0.17$ & $1.80\pm0.11$ & $1.11\pm0.10$ & $\nu>max\{\nu_c^f,\nu_m^f\}$ & (2) \\
  061007(o) & 1.26&$3.93\pm0.03$  & $-3.01\pm0.54$ & $0.76\pm0.03$ & $2.56\pm0.04$ & $0.76\pm0.03$ & $\nu_m^f<\nu<\nu_c^f$        & (1) \\
  070103(x) & 2.62&$0.55\pm0.16$  & $-1.32\pm0.05$ & $1.31\pm0.14$ & $2.43\pm0.06$ & $1.10\pm0.27$ & $\nu>max\{\nu_c^f,\nu_m^f\}$ & (3) \\
  070208(x) & 1.17&$1.09\pm0.25$  & $-1.29\pm0.06$ & $0.83\pm0.23$ & $2.39\pm0.08$ & $1.05\pm0.19$ & $\nu>max\{\nu_c^f,\nu_m^f\}$ & (3) \\
  070318(o) & 0.84&$0.54\pm0.11$  & $-1.10\pm0.04$ & $1.38\pm0.06$ & $2.11\pm0.06$ &   0.78        & $\nu_m^f<\nu<\nu_c^f$        & (2) \\
  070411(o) & 2.95&$0.40\pm0.01$  & $-1.25\pm0.00$ & $1.43\pm0.01$ & $2.30\pm0.00$ & $0.75\pm0.10$ & $\nu_m^f<\nu<\nu_c^f$        & (2) \\
  070419A(o)& 0.97&$1.09\pm0.09$  & $-1.20\pm0.02$ & $1.04\pm0.05$ & $2.37\pm0.03$ & $0.83\pm0.16$ & $\nu_m^f<\nu<\nu_c^f$        & (2) \\
  071010A(o)& 0.98&$2.36\pm0.43$  & $-0.74\pm0.01$ & $0.37\pm0.25$ & $1.92\pm0.05$ & $0.76\pm0.23$ & $\nu_m^f<\nu<\nu_c^f$        & (2) \\
  071031(o) & 2.05&$0.63\pm0.00$  & $-0.85\pm0.00$ & $1.40\pm0.00$ & $1.79\pm0.00$ & $0.90\pm0.10$ & $\nu>max\{\nu_c^f,\nu_m^f\}$ & (2) \\
  080330(o) & 1.51&$0.34\pm0.03$  & $-1.77\pm0.12$ & $1.32\pm0.05$ & $3.03\pm0.16$ &   0.99        & $\nu_m^f<\nu<\nu_c^f$        & (2) \\
  080710(o) & 0.85&$1.08\pm0.01$  & $-1.00\pm0.01$ & $0.92\pm0.00$ & $2.00\pm0.00$ & $1.00\pm0.02$ & $\nu>max\{\nu_c^f,\nu_m^f\}$ & (2) \\
  080810(o) & 3.35&$1.34\pm0.05$  & $-1.20\pm0.00$ & $0.90\pm0.03$ & $2.41\pm0.01$ & $0.51\pm0.02$ & $\nu_m^f<\nu<\nu_c^f$        & (2) \\
  081007(o) & 0.53&$3.01\pm0.07$  & $-4.17\pm0.30$ & $1.16\pm0.07$ & $2.72\pm0.14$ & $0.86\pm0.07$ & $\nu_m^f<\nu<\nu_c^f$        & (1) \\
  081008(o) & 1.97&$3.38\pm0.12$  & $-2.84\pm0.34$ & $0.41\pm0.15$ & $2.20\pm0.08$ & $1.10\pm0.05$ & $\nu>max\{\nu_c^f,\nu_m^f\}$ & (1) \\
  081203A(o)& 2.05&$2.20\pm0.01$  & $-1.49\pm0.01$ & $0.40\pm0.01$ & $2.91\pm0.01$ & $0.90\pm0.01$ & $\nu_m^f<\nu<\nu_c^f$        & (2) \\
  090102(o) & 1.55&$3.21\pm0.40$  & $-5.15\pm2.50$ & $1.15\pm0.21$ & $2.86\pm0.36$ & $0.93\pm0.18$ & $\nu_m^f<\nu<\nu_c^f$        & (1) \\
  090313(o) & 3.38&$1.23\pm0.10$  & $-1.25\pm0.03$ & $0.71\pm0.09$ & $2.33\pm0.04$ &   1.2         & $\nu>max\{\nu_c^f,\nu_m^f\}$ & (2) \\
  100901A(x)& 1.41&$0.80\pm0.07$  & $-1.45\pm0.03$ & $1.04\pm0.06$ & $2.60\pm0.04$ & $1.11\pm0.06$ & $\nu>max\{\nu_c^f,\nu_m^f\}$ & (3) \\
  100906A(o)& 1.73&$1.87\pm0.31$  & $-1.00\pm0.01$ & $0.63\pm0.17$ & $2.21\pm0.14$ & $1.00\pm0.10$ & $\nu_m^f<\nu<\nu_c^f$        & (3) \\
  110205A(o)& 2.22&$3.88\pm0.08$  & $-5.17\pm0.24$ & $0.86\pm0.06$ & $2.68\pm0.08$ & $0.84\pm0.04$ & $\nu_m^f<\nu<\nu_c^f$        & (1) \\
  110213A(o)& 1.46&$1.54\pm0.08$  & $-0.91\pm0.02$ & $0.83\pm0.04$ & $2.04\pm0.03$ & $1.12\pm0.12$ & $\nu_m^f<\nu<\nu_c^f$        & (2) \\
  120118B(x)& 2.94&$0.79\pm0.20$  & $-0.91\pm0.08$ & $1.25\pm0.21$ & $1.88\pm0.11$ & $0.98\pm0.15$ & $\nu>max\{\nu_c^f,\nu_m^f\}$ & (3) \\
  120224A(x)& 1.10&$0.80\pm0.11$  & $-1.00\pm0.03$ & $1.20\pm0.11$ & $2.00\pm0.03$ & $1.10\pm0.13$ & $\nu>max\{\nu_c^f,\nu_m^f\}$ & (3) \\
  121209A(x)& 2.10&$0.92\pm0.21$  & $-1.23\pm0.03$ & $1.00\pm0.20$ & $2.31\pm0.05$ & $1.11\pm0.13$ & $\nu>max\{\nu_c^f,\nu_m^f\}$ & (3) \\
  130427A(o)& 0.33&$1.63\pm0.10$  & $-1.67\pm0.07$ & $1.48\pm0.06$ & $2.40\pm0.10$ & $0.70\pm0.05$ & $\nu_m^f<\nu<\nu_c^f$        & (1) \\
  140512A(o)& 0.73&$3.04\pm0.09$  & $-1.93\pm0.07$ & $1.15\pm0.03$ & $2.72\pm0.02$ & $0.86\pm0.01$ & $\nu_m^f<\nu<\nu_c^f$        & (1) \\
  151112A(x)& 4.10&$1.57\pm0.56$  & $-1.04\pm0.04$ & $0.42\pm0.55$ & $2.05\pm0.05$ & $1.16\pm0.16$ & $\nu>max\{\nu_c^f,\nu_m^f\}$ & (2)(5) \\
  161023A(o)& 2.71&$5.44\pm0.13$  & $-1.06\pm0.01$ & $0.01\pm0.19$ & $2.32\pm0.16$ & $0.66\pm0.08$ & $\nu_m^f<\nu<\nu_c^f$        & (1)(6) \\
  \hline
    \end{tabular*}
    \begin{tablenotes}
        \footnotesize
        \item References. (1) \citealp{2020ApJ...895...94Y}; (2) \citealp{2013ApJ...776..120Y}; (3) \citealp{2022ApJ...925...54T}; (4) \citealp{2023Sci...380.1390L}; (5) \citealp{2014Sci...343...38V}; (6) \citealp{2016ApJ...833..100H};
      \end{tablenotes}
  \end{threeparttable}
    }
  \label{tab:addlabel}%
  \end{center}
\end{table*}
\section{The correlations of $\Gamma_0$ with $E_{\rm \gamma,iso}$, $L_{\rm \gamma,iso}$ and $\ E_{p,z}$}
\label{Sec: 4}

\subsection{$ \Gamma_0$-$ E_{\rm \gamma,iso} $ and $ \Gamma_0$-$ L_{\rm \gamma,iso}$ correlation}
\label{Sec: 4.1}
Using $ \Gamma_0$ calculated in Section 3, we can further explore its correlations with other relevant parameters, e.g., $E_{\rm \gamma,iso}$, $L_{\rm \gamma,iso}$ and$\ E_{\rm p,z}$. By collecting a set of GRBs (from Swift/XRT), which are characterized by an onset bump feature in afterglow light curves, \cite{2010ApJ...725.2209L} found a tight correlation between $ \Gamma_0$ and $ E_{\rm \gamma,iso}$ for a constant-density medium with the radiative efficiency $\eta$ =0.2, $ \Gamma_0$$\simeq$ 195 $E^{0.27}_{\rm \gamma,iso,52}$. This tight relation not only suggests the $E_{\rm \gamma,iso}$ is a good indicator of $ \Gamma_0$, but also provides a new approach for determining the physical properties of the prompt emission of GRBs. In addition, \cite{2013ApJ...774...13L} found that the correlation can be expressed as $ \Gamma_0$$\simeq$ 26.92 $E^{0.26}_{\rm \gamma,iso,50}$ using the GRBs with deceleration feature. In this work, we will also find a similar slope for this correlation.

In particular, the correlation between $\Gamma_0$ and $E_{\rm \gamma,iso}$ can be written as
\begin{equation}
\log \Gamma_0=a+b \times \log \left(\frac{E_{\gamma, \text { iso }}}{10^{52} \mathrm{erg}}\right)
\label{Eq: E3}
\end{equation}

Then, by employing the likelihood function we can obtain the best fitting values of $a, b$, and the intrinsic scatter $\sigma_{\rm int}$. It is worth noting that the smaller the value of $\sigma_{\rm int}$, the higher the agreement between the model predictions and the data points, indicating a stronger correlation. The likelihood function can be written as \citep{2005physics..11182D}

\begin{equation}\
\begin{split}
\mathcal{L}\left(a, b, \sigma_{\mathrm{int}}\right) \propto \prod_i \frac{1}{\sqrt{\sigma_{\mathrm{int}}^2+\sigma_{y_i}^2+b^2 \sigma_{x_i}^2}}\times \\
\exp \left[-\frac{y_i-a-b x_i}{2\left(\sigma_{\mathrm{int}}^2+\sigma_{y_i}^2+b^2 \sigma_{x_i}^2\right)}\right]
\end{split}
\label{Eq: E4}
\end{equation}
Here we define $y$ = {\rm log}$ \Gamma _{0}$, $x$ = log ($E_{\rm \gamma ,iso}$/{$10^{52}$erg}).
The maximization of the likelihood function and the determination of the parameter error both are performed employing the Markov Chain Monte Carlo (MCMC) algorithm with the public \texttt{EMCEE} package (\citealp{2013PASP..125..306F}). Similar to the results in \cite{2010ApJ...725.2209L} and \cite{2013ApJ...774...13L}, we also found a correlation, and the results are shown in Figure \ref{Fi:E1}. The best fitting results for $\eta=0.2$ are $a$ = 1.95$\pm$0.04, $b$ = 0.24$\pm$0.03 and $\sigma_{\rm int}$ = 0.20$\pm$0.03. Therefore, the correlation can be expressed as

\begin{equation}
\log \Gamma_0=(1.95 \pm 0.04)+(0.24 \pm 0.03) \log E_{\gamma,\text { iso },52}
\label{Eq: E5}
\end{equation}
or $ \Gamma_0$$\simeq$$E^{0.24\pm0.03}_{\rm \gamma,iso,52}$, which is consistent with the results of previous works.
Consider the correlation expression in \cite{2010ApJ...725.2209L} as $ \Gamma_0$$\simeq$ 195 $E^{0.27}_{\rm \gamma,iso,52}$, the slope and intercept in our result are smaller than it. This slight discrepancy could be attributed to the utilization of a distinct circumburst medium instead of the homogeneous interstellar ambient.

The correlation between $E_{\rm \gamma,iso}$ and $\Gamma_0$ is crucial for elucidating the Amati relation (\citealp{2002A&A...390...81A}) ($E_{\rm p}$ - $E_{\rm \gamma,iso}$) across various theoretical models of GRBs. Furthermore, \cite{2012ApJ...751...49L} found the more tighter correlation between the mean isotropic luminosity $L_{\rm \gamma,iso}$ and $\Gamma_0$, $ \Gamma_0$$\simeq$249$L^{0.30}_{\rm \gamma,iso,52}$, which may be more intrinsic than the $E_{\rm \gamma,iso}$-$\Gamma_0$ correlation. In this study, we also confirmed the correlation of $L_{\rm \gamma,iso}$ and $\Gamma_0$ using our sample dataset. The mean isotropic luminosity $L_{\rm \gamma,iso}$ can be written as $L_{\rm \gamma,iso}$ = $E_{\rm \gamma,iso}$/$T_{\rm 90,z}$, and the corresponding results for 34 GRBs are reported in Table \ref{tab:addlabe2}. Applying the same methodological steps as mentioned above with $y$ = log$\Gamma _{0}$ and $x$ = log($L_{\rm \gamma ,iso}$/{$10^{49}$erg $s^{-1}$}) for equation \ref{Eq: E4}, then we obtain the fitting result for $\eta$ = 0.2 are $a$ = 1.66$\pm$0.10, $b$ = 0.20$\pm$0.05 and $\sigma_{\rm int}$ = 0.23$\pm$0.04. From this, the correlation can be expressed as
\begin{equation}
\log \Gamma _0 = (1.66\pm0.10) + (0.20 \pm 0.05) \log L_{\gamma ,\text {iso},49 }
\label{Eq: E6}
\end{equation}
or $ \Gamma_0$$\simeq$$L^{0.20\pm0.05}_{\rm \gamma,iso,49}$.

The above results are also depicted on the right side of Figure \ref{Fi:E1}. Similar to the correlation of $E_{\rm \gamma,iso}$ - $\Gamma_0$, the slope and coefficient of the $L_{\rm \gamma,iso}$ - $\Gamma_0$ are smaller than that of \cite{2012ApJ...751...49L}. Furthermore, for the $L_{\rm \gamma,iso}$ - $\Gamma_0$ correlation, the value of $\sigma_{\rm int}$ is slightly larger than for the $E_{\rm \gamma,iso}$ - $\Gamma_0$ correlation, which means the latter is a bit more compact. The result is contrary to the finding of \cite{2012ApJ...751...49L}. This suggests that for different circumburst mediums, the degree of $L_{\rm \gamma,iso}$ dependence on $\Gamma_0$ is substantially weaker than that of $E_{\rm \gamma,iso}$ dependence on $\Gamma_0$.

\subsection{$ E_{\rm \gamma,iso}$ - $\ E_{\rm p,z}$ - $ \Gamma_0$ and $L_{\rm \gamma,iso}$ - $\ E_{\rm p,z}$ - $ \Gamma_0$ correlation}
\label{Sec: 4.2}
Given that $\ E_{\rm p,z}$ depends not only on $ E_{\rm \gamma,iso}$ ($L_{\gamma,iso}$) but also on the initial Lorentz factor $ \Gamma_0$, \cite{2015ApJ...813..116L} tried to explore the multivariate correlations associated with $ \Gamma_0$ using the linear regression method. It is interesting to note that among the six sets of multivariate correlations, $ E_{\rm \gamma,iso}$ - $ \Gamma_0$ - $\ E_{\rm p,z}$ and $ L_{iso}$ - $ \Gamma_0$ - $\ E_{\rm p,z}$ correlations are much tighter than the correlations $ E_{\rm \gamma,iso}$ - $ \Gamma_0$ and $ L_{iso}$ - $ \Gamma_0$ (e.g., Table 2 of \citealp{2015ApJ...813..116L}). The best fitting results can be expressed as $E_{\rm \gamma,iso,52}$ $\propto$ $\Gamma^{0.96\pm0.20}_0$$\ E^{0.98\pm0.19}_{\rm p,z}$ and $L_{iso,46}$ $\propto$ $\Gamma^{1.32\pm0.19}_0$$\ E^{1.34\pm0.14}_{\rm p,z}$ ($k$ = 0 and $\eta$ = 0.2), respectively.

In this paper, we also quantitatively investigate the correlations of $ E_{\rm \gamma,iso}$ - $\ E_{\rm p,z}$ - $ \Gamma_0$. Be different the $ L_{iso}$ - $ \Gamma_0$ - $\ E_{\rm p,z}$ correlation in \cite{2015ApJ...813..116L}, we investigate the correlation between the isotropic equivalent peak luminosity $L_{\rm iso}$, $\ E_{\rm p,z}$ and $ \Gamma_0$. For each GRB in our sample, we collected $\ E_{\rm p}$ in the literatures and calculated $\ E_{\rm p,z}$ = $\ E_{\rm p}$ (1+$z$). Following the same approach as in the Section \ref{Sec: 4.1}, the multi-variable correlation $ E_{\rm \gamma,iso}$ - $\ E_{\rm p,z}$ - $ \Gamma_0$ can be written as follows,

\begin{equation}
\log \left(\frac{E_{\gamma, \text { iso }}}{10^{52} \mathrm{erg}}\right)=a+b \times \log \Gamma_0+c \times \log \left(\frac{E_{\mathrm{p}, \mathrm{z}}}{\mathrm{keV}}\right).
\label{Eq: E7}
\end{equation}

We then write the corresponding likelihood function as follows,
\begin{equation}
\begin{split}
\mathcal{L}\left(a, b, c, \sigma_{\text {int }}\right) \propto \prod_i \frac{1}{\sqrt{\sigma_{\mathrm{int}}^2+\sigma_{y_i}^2+b^2 \sigma_{x_{1, i}}+c^2 \sigma_{x_{2, i}}}}\\
 \times \exp \left[-\frac{y_i-a-b x_{1, i}-c x_{2, i}}{2\left(\sigma_{\text {int }}^2+\sigma_{y_i}^2+b^2 \sigma_{x_{1, i}}^2+c^2 \sigma_{x_{2, i}}^2\right)}\right].
\end{split}
\label{Eq: E8}
\end{equation}
Here we take $y$ = ($E_{\rm \gamma ,iso}$/{$10^{52}${\rm erg}), $x_1$ = log$\Gamma_{0}$, and $x_2$ = ($E_{\rm p,z}$/{\rm keV}). The coefficients from the best fitting with $\eta$ = 0.2 are $a$ = -4.39$\pm$0.65, $b$ = 1.36$\pm$0.39, $c$ = 0.82$\pm$0.20 and $\sigma_{\rm int}$ = 0.52$\pm$0.07, respectively. Therefore, the correlation of $ E_{\rm \gamma,iso}$ - $\ E_{\rm p,z}$ - $ \Gamma_0$ can be expressed as
\begin{equation}
\begin{split}
{\rm log} E_{\rm \gamma ,iso} =(-4.39\pm0.65)+(1.36\pm0.39){\rm log} \Gamma _{0}\\
 +(0.82\pm0.20){\rm log}(E_{\rm p,z}/{\rm keV}),
\end{split}
\label{Eq: E9}
\end{equation}
or $E_{\rm \gamma ,iso,52}$$\simeq$$ \Gamma^{1.36\pm0.39}_0$$E^{0.82\pm0.20}_{\rm p,z}$. Unlike \cite{2015ApJ...813..116L}, the compactness of the $ E_{\rm \gamma,iso}$-$\ E_{\rm p,z}$-$ \Gamma_0$ is not significantly improved compared with the $ E_{\rm \gamma,iso}$-$ \Gamma_0$. This could be due to the large dispersion of $ E_{\rm \gamma,iso}$ and $\ E_{\rm p,z}$ correlations in the current sample.

Furthermore, we explore the $ L_{\rm \gamma,iso}$-$\ E_{\rm p,z}$-$ \Gamma_0$ correlation by setting $y$ = ($L_{\rm \gamma ,iso}$/{$10^{49}$\rm erg/s), $x_1$ = log$\Gamma_{0}$, and $x_2$ = ($E_{\rm p,z}$/\rm keV) for equation \ref{Eq: E8}, and the equation can be written as follows

\begin{equation}
\log \left(\frac{L_{\gamma, \text { iso }}}{10^{49} \mathrm{erg} / \mathrm{s}}\right)=a+b \times \log \Gamma_0+c \times \log \left(\frac{E_{\mathrm{p}, \mathrm{z}}}{\mathrm{keV}}\right).
\label{Eq: E10}
\end{equation}

The corresponding coefficients from the best fitting are found to be $a$ = -1.91$\pm$0.83, $b$ = 1.05$\pm$0.50, $c$ = 0.71$\pm$0.25 and $\sigma_{\rm int}$ = 0.66$\pm$0.09 for $\eta$ = 0.2, i.e.,
\begin{equation}
\begin{split}
\log L_{\gamma, \text { iso }}=(-1.91 \pm 0.83)+(1.05 \pm 0.50) \log \Gamma_0\\
+(0.66 \pm 0.09) \log \left(E_{\mathrm{p}, \mathrm{z}} / {\rm keV}\right)
\end{split}
\label{Eq:E11}
\end{equation}
or $L_{\rm \gamma ,iso,52}$$\simeq$$ \Gamma^{1.05\pm0.50}_0$$E^{0.66\pm0.09}_{\rm p,z}$. We present the fitting results in Figure \ref{F2:E2}.

It is worth noting that the intrinsic scatter $\sigma_{\rm int}$ of $L_{\rm \gamma,iso}$-$\ E_{\rm p,z}$-$ \Gamma_0$ is larger than that of $E_{\rm \gamma,iso}$-$\ E_{\rm p,z}$-$ \Gamma_0$, implying no enhancement in the compactness of correlation. This phenomenon aligns with the result of our two-parameter, indicating that the correlation between $L_{\rm \gamma,iso}$ and $ \Gamma_0$ is weaker than that of $E_{\rm \gamma,iso}$ and $ \Gamma_0$ for the current sample. As previously mentioned, the reason for this phenomenon may be due to different circumburst medium conditions that alter the dependence of $L_{\rm \gamma,iso}$ on $ \Gamma_0$.

\begin{figure*}
	\centering
	\includegraphics[width=0.4\textwidth, angle=0]{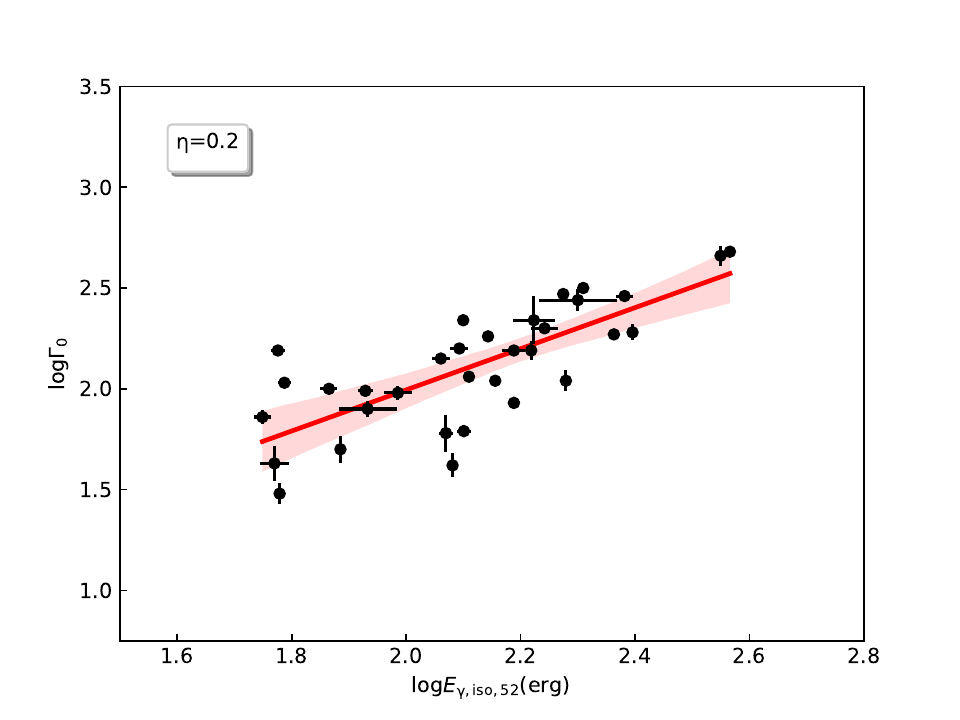}	\includegraphics[width=0.4\textwidth, angle=0]{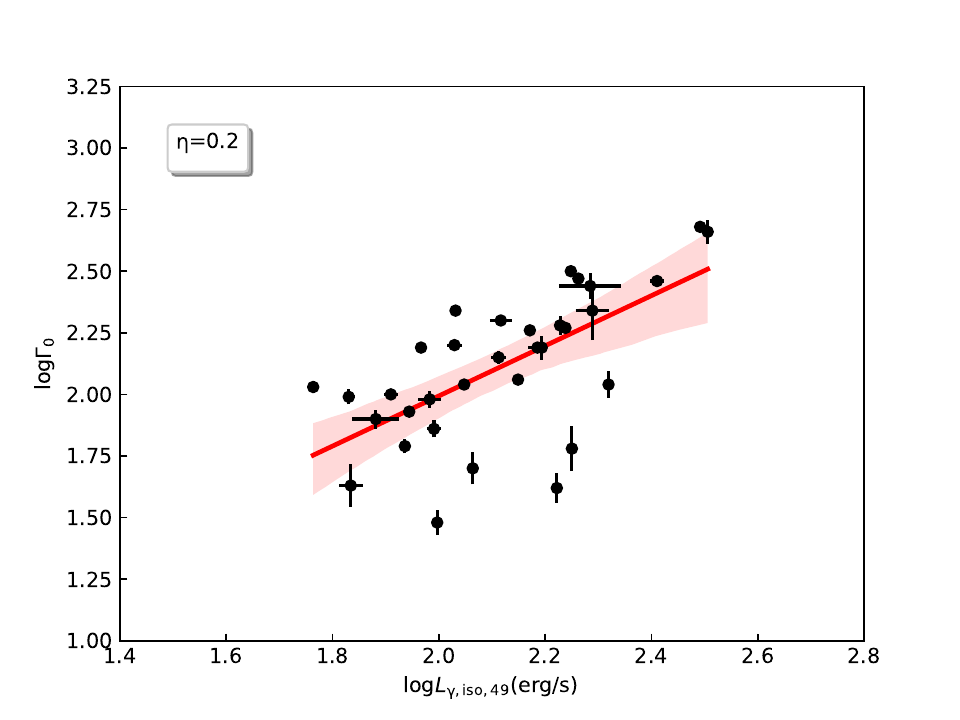}
	\caption{Correlations between $ \Gamma_0$ and $ E_{\rm \gamma,iso}$ ($ L_{\rm \gamma,iso}$). }
	\label{F2:E2}
\end{figure*}

\begin{figure*}
	\centering
	\includegraphics[width=0.4\textwidth, angle=0]{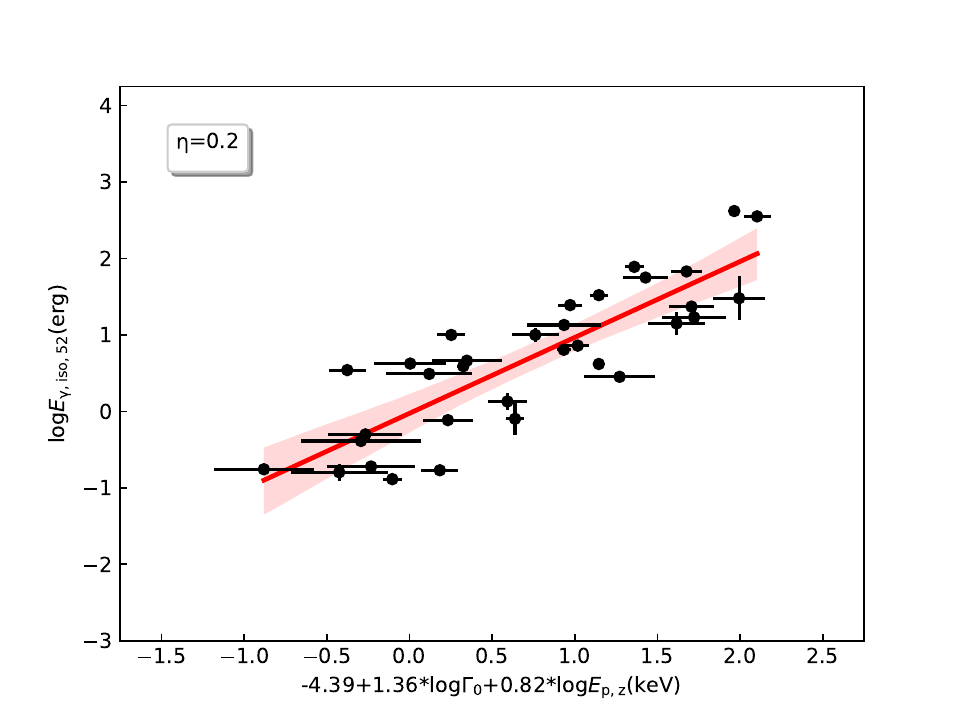}	\includegraphics[width=0.4\textwidth, angle=0]{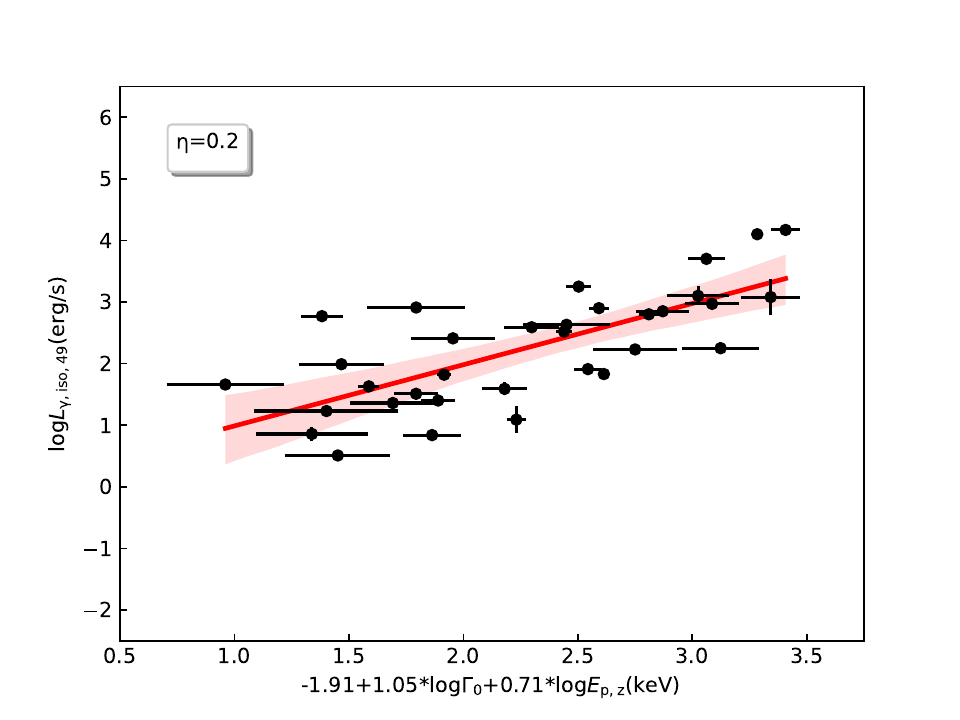}
	\caption{Three-parameter correlations: $ E_{\rm \gamma,iso}$ - $ \Gamma_0$ - $E_{\rm p,z}$(left) and $ L_{\rm \gamma,iso}$ - $ \Gamma_0$ - $E_{\rm p,z}$(right). }
	\label{Fi:E1}
\end{figure*}

\section{Discussion and Conclusion}
\label{Sec: conclusions}
In this work, we infer the initial Lorentz factor for different circumburst mediums. We first select a sample of 33 GRBs with redshift measurements, which contains 7 X-ray band samples and 26 optical band samples. For the circumburst medium density, $n$ =$n_0$$(R/R_0)^{-k}$, we calculate the initial Lorentz factor $ \Gamma_0$ for the 33 GRB samples in various circumburst mediums, finding a varied $ \Gamma_0$ from 50 to 500 for different radiative efficiencies considered by us.

We conduct a comprehensive analysis of the linear relationship among the initial Lorentz factor $ \Gamma_0$ and the crucial observational parameters of the GRB afterglows, i.e., $E_{\rm \gamma,iso}$, $L_{\rm\gamma,iso}$ and $\ E_{\rm p,z}$. We find a tight correlation between the $ \Gamma_0$ and $E_{\rm \gamma,iso}$ ($L_{\rm\gamma,iso}$), which can be expressed as $\Gamma_0$ $\propto$ $E^{0.24}_{\rm \gamma,iso,52}$ ($\Gamma_0$ $\propto$ $L^{0.20}_{\rm \gamma,iso.49}$) for $\eta$=0.2. These correlations are consistent with previous studies (\citealp{2010ApJ...725.2209L}; \citealp{2013ApJ...774...13L}). It is worth noting that the compactness of the $L_{\rm\gamma,iso}$ - $ \Gamma_0$ correlation for the current sample is slightly weaker than that of the $E_{\rm\gamma,iso}$ - $ \Gamma_0$ correlation. This could be attributed to the less interdependence between $L_{\rm\gamma,iso}$ and $ \Gamma_0$ for distinct circumburst medium.

Furthermore, we verify the correlations of $E_{\rm \gamma,iso}$ - $ \Gamma_0$ - $\ E_{\rm p,z}$ and $L_{\rm \gamma,iso}$ - $ \Gamma_0$-$\ E_{\rm p,z}$, i.e., $E_{\rm \gamma,iso,52}$ $\propto$ $\Gamma^{1.36}_0$$E^{0.82}_{\rm p,z}$ ($L_{\rm \gamma,iso,49}$ $\propto$ $\Gamma^{1.05}_0$$E^{0.66}_{\rm p,z}$). The compactness of the correlations is  more dispersed than that of between the two parameters, slightly different from the findings in \cite{2015ApJ...813..116L}. Interestingly, it is found that the $L_{\rm\gamma,iso}$-$ \Gamma_0$-$\ E_{\rm p,z}$ correlation is a bit more discrete than the $E_{\rm\gamma,iso}$-$ \Gamma_0$-$\ E_{\rm p,z}$ correlation. This finding is consistent with the results from pair correlations, suggesting that the dependence of $L_{\rm \gamma,iso}$ on $ \Gamma_0$ is weaker for our sample for different circumburst mediums. These correlations are crucial in understanding the circumurst medium GRBs' origin.

\begin{table*}[htbp]
  \small
  \begin{center}
  \caption{The fireball radius ($R_0$), duration ($T_{90}$), peak time ($t_{\rm p}$), isotropic luminosity ($L_{\rm\gamma,iso}$), isotropic energy ($E_{\rm \gamma,iso}$) and the peak energy of the $\upsilon$$f_{\upsilon}$ spectrum ($E_{\rm p}$) of prompt gamma-rays, as well as the initial Lorentz factor ($\Gamma_0$) for GRBs in our sample.}{
    \begin{threeparttable}
    \begin{tabular*}{1.0\linewidth}{c|c|c|c|c|c|c|c|c}
  \hline
  GRB&  $R_0$            &  $T_{90}$ & $\ t_{\rm p}$& $E_{\rm \gamma,iso}$ &$L_{\rm\gamma,iso}$& $\ E_{\rm p}$&$ \Gamma_0$  &       Refs.               \\
     & ($10^{18}\rm cm$) &(s)        &       (s)    &    ($10^{52}\rm erg$)&($10^{49}\rm erg/s$)&   (keV)     &             &                           \\
  \hline
  990123(o)   & 0.90 & 63   & 47$\pm$10        & 356$\pm$7      & 14748.57$\pm$290  & 1334$\pm$53.6   & 458.08$\pm$52.22  & (1)(2)(3)    \\
  041219A(o)  & 0.09 & 520  & 1331.3$\pm$3.4   & 10$\pm$1       & 25.19$\pm$2.52    & 196$\pm$97      & 84.88$\pm$2.64    &  (4)(5)  \\
  050730(o)   & 0.30 & 154.6& 590.7$\pm$131.5  & 13.37$\pm$1.22 & 429.73$\pm$39.21  &156.52$\pm$69.48 & 156.2$\pm$17.41   & (2)(6)(7)(8) \\
  060605(o)   & 0.10 & 79.84& 590$\pm$45       & 2.83$\pm$0.45  & 169.18$\pm$26.9   & 490$\pm$251     & 140.12$\pm$7.93   & (1)(6)(9)(10)     \\
  060607A(o)  & 0.07 &103.03& 179$\pm$3        & 23.42$\pm$1.49 & 927.89$\pm$59.03  & 575$\pm$200     & 292.12$\pm$6.34   & (1)(9)(10)       \\
  060904B(o)  & 0.05 &189.98& 467.9$\pm$48.4   & 0.77$\pm$0.1   & 6.90$\pm$0.90     & 135$\pm$42      & 97.01$\pm$6.71    & (1)(6)(7)(9)       \\
  061007(o)   & 0.10 & 75.74& 77$\pm$1         & 421$\pm$41.9   &12567.7$\pm$1250.8 & 965$\pm$27      & 480.15$\pm$11.52  & (1)(9)(10)(11)       \\
  070103(X)   & 1.00 & 18.6 & 685.5$\pm$64.3   & 0.50$\pm$0.03  & 97.33$\pm$5.84    &47.13$\pm$17.93  & 50.05$\pm$7.56    & (7)(12)(13)       \\
  070208(X)   & 1.00 & 48   & 968.1$\pm$72.9   & 0.16$\pm$0.04  & 7.22$\pm$1.80     & 65.85$\pm$32.9  & 42.91$\pm$8.5     & (7)(8)(12)       \\
  070318(o)   & 0.09 & 63   & 507$\pm$46       & 1.35$\pm$0.33  & 39.34$\pm$9.62    & 359.75$\pm$72.07& 95.87$\pm$7.34    & (2)(6)(10)(14)       \\
  070411(o)   & 0.10 & 101  & 739$\pm$10       & 10$\pm$2       & 391.49$\pm$78.29  & 120.38$\pm$39.04& 154.6$\pm$6.74    & (2)(6)(8)(10)       \\
  070419A(o)  & 0.004& 116  & 765$\pm$30       & 0.19$\pm$0.02  & 3.23$\pm$0.34     & 27$\pm$17.6     & 107.2$\pm$6.5     & (1)(2)(6)(10)       \\
  071010A(o)  & 0.03 & 6    & 586$\pm$66       & 0.13$\pm$0.02  & 42.9$\pm$6.6      & 73.96$\pm$2.30  & 72.97$\pm$5.82    & (2)(6)(10)(14)       \\
  071031(o)   & 0.02 &180   & 1213$\pm$2       & 3.9$\pm$0.6    & 66.08$\pm$10.17   &43.95$\pm$2.30   & 157.44$\pm$4.74   & (2)(6)(10)(14)       \\
  080330(o)   & 0.04 &60.36 & 578$\pm$25       & 0.41$\pm$0.06  & 17.05$\pm$2.50    & 20$\pm$19       & 99.4$\pm$4.36     & (6)(8)(9)(10)       \\
  080710(o)   & 0.01 &120   & 1934$\pm$46      & 0.8$\pm$0.4    & 12.3$\pm$6.15     &553.35$\pm$2.30  & 78.8$\pm$7.03     & (2)(6)(10)(14)       \\
  080810(o)   &0.50  &108   & 117$\pm$2        & 30$\pm$20      & 12083.3$\pm$805.56& 1364$\pm$320    & 274.4$\pm$33.85   & (1)(2)(6)(10)       \\
  081007(o)   & 0.01 &8     & 123.5$\pm$0      & 0.17$\pm$0.02  & 32.51$\pm$3.83    & 61$\pm$15       & 156.25$\pm$6.69   & (2)(11)      \\
  081008(o)   & 0.025&185   & 163$\pm$2        & 4.17$\pm$0.19  & 66.88$\pm$3.05    & 267.3$\pm$2.3   & 220.03$\pm$6.96   & (1)(2)(10)(14)(16)       \\
  081203A(o)  & 0.10 & 294  & 295$\pm$2        & 17$\pm$4       & 176.36$\pm$41.50  & 1541$\pm$757    & 201.54$\pm$7.15   & (1)(2)(6)(10)       \\
  090102(o)   & 1.70 & 28.32& 50$\pm$1         & 14$\pm$5       & 1259.11$\pm$496.82& 1174$\pm$38     & 189.6$\pm$61.01   & (1)(9)(10)(11)       \\
  090313(o)   & 0.08 & 78   & 1315$\pm$109     & 4.6$\pm$0.5    & 258.01$\pm$28.04  & 55.65$\pm$28.37 & 115.25$\pm$5.94   & (2)(6)(10)(11)       \\
  100901A(X)  & 1.00 & 439  & 1260$\pm$76      & 4.22$\pm$0.5   & 23.15$\pm$2.74    & 107.88$\pm$53.74& 62.16$\pm$4.24    & (10)(12)(13)(16)       \\
  100906A(o)  & 0.10 &114   & 101$\pm$4        & 33.4$\pm$3     & 798.96$\pm$3.83   & 158$\pm$16      & 314.57$\pm$9.73   & (1)(2)(6)(10)       \\
  110205A(o)  & 0.05 & 257  & 948$\pm$3        & 56$\pm$6       & 701.63$\pm$75.18  & 715$\pm$239     & 186.25$\pm$5.02   & (1)(9)(10)       \\
  110213A(o)  & 0.04 & 48   & 293$\pm$12       & 6.4$\pm$0.6    & 328$\pm$30.75     & 241$\pm$13      & 184.04$\pm$6.63   & (1)(6)(9)(10)       \\
  120118B(X)  & 1.00 & 23.26& 4786.3$\pm$330.63& 3.47$\pm$0.16  & 588.22$\pm$27.12  & 43.1$\pm$3.9    & 41.27$\pm$5.62    & (12)(13)(15)(17)       \\
  120224A(X)  & 1.00 & 8.13 & 1379$\pm$121     & 0.175$\pm$0.01 & 45.20$\pm$2.58    & 33.29$\pm$21.8  & 30.23$\pm$3.54    & (12)(17)       \\
  121209A(X)  & 1.00 & 42.7 & 752$\pm$81       & 24.31$\pm$0.84 & 1764.9$\pm$60.98  & 494$\pm$0       & 109.54$\pm$13.37  & (4)(12)(18)       \\
  130427A(o)  & 4.00 & 163  & 13.5$\pm$0.01    & 77.01$\pm$7.88 & 630.25$\pm$64.49  & 1371.3$\pm$10.7 & 189.6$\pm$17.24   & (2)(18)       \\
  140512A(o)  & 0.80 & 155  & 209.3$\pm$1.93   & 7.25$\pm$0.61  & 80.69$\pm$6.79    & 1011$\pm$145    & 108.89$\pm$3.6    &  (2)(13)       \\
  151112A(X)  & 1.00 & 19.32& 5118$\pm$382     & 3.09$\pm$0.36  & 815.68$\pm$95.03  & 73.25$\pm$28.15 & 59.77$\pm$12.54   &  (12)(13)       \\
  161023A(o)  & 0.09 & 49.98& 211$\pm$4        & 68.17$\pm$9.74 & 5057.51$\pm$722.61& 604.4$\pm$121.66& 289.29$\pm$10.97  &  (19)       \\
  \hline
    \end{tabular*}
    \begin{tablenotes}
        \footnotesize
        \item References. (1)\citealp{2015ApJ...813..116L}; (2)\citealp{2021ApJ...908..242D}; (3)\citealp{2012MNRAS.420..483G}; (4)\citealp{2018ApJ...852...53R}; (5)\citealp{2023A&A...675A.175B}; (6)\citealp{2013ApJ...776..120Y}; (7)\citealp{2004ApJ...606L..29L}; (8)\citealp{2018ApJ...863...50S} (9)\citealp{2020ApJS..248...21H}; (10)\citealp{2013ApJ...774...13L}; (11)\citealp{2012MNRAS.421.1256N}; (12)\citealp{2022ApJ...925...54T}; (13)\citealp{2019ApJ...883...97Z}; (14)\citealp{2004ApJ...616..331G}; (15)\citealp{2021MNRAS.507.1047Y}; (16)\citealp{2013ApJ...774...13L}; (17)\citealp{2021MNRAS.508...52L}; (18)\citealp{2016A&A...587A..40P}; (19)\citealp{2019ApJ...876...77X};
      \end{tablenotes}
  \end{threeparttable}
    }
  \label{tab:addlabe2}%
\end{center}

\end{table*}%

\section*{Acknowledgements}
We thank the anonymous referee for thoughtful comments. This work is supported by the National Natural Science Foundation of China (Grant No. U2038106), the Natural Science Foundation of Jiangxi Province of China (grant No. 20242BAB26012), Shandong Provincial Natural Science Foundation (ZR2021MA021) and and Manned Spaced Project (CMS-CSST-2021-A12).





\section*{References}
\newcommand{\arnps}{Annu.\ Rev.\ Nucl.\ Part.\ Sci.}
\newcommand{\al}{Astron.\ Lett.}
\newcommand{\aap}{Astron.\ Astrophys.}
\newcommand{\apj}{Astrophys.\ J.}
\newcommand{\apjl}{Astrophys.\ J.}
\newcommand{\apjs}{Astrophys.\ J.\ Suppl.\ Ser.}
\newcommand{\epja}{Eur.\ Phys.\ J. A}
\newcommand{\iaucirc}{IAU Circ.}
\newcommand{\jaa}{J.\ Astron.\ Astrophys.}
\newcommand{\nat}{Nature}
\newcommand{\mnras}{Mon.\ Not.\ R.\ Astron.\ Soc.}
\newcommand{\npa}{Nucl.\ Phys. A}
\newcommand{\physrep}{Phys.\ Rep.}
\newcommand{\prc}{Phys.\ Rev. C}
\newcommand{\prd}{Phys.\ Rev. D}
\newcommand{\prl}{Phys.\ Rev.\ Lett.}
\newcommand{\ptp}{Prog.\ Theor.\ Phys.}
\newcommand{\ppnp}{Prog.\ Part.\ Nucl.\ Phys.}
\newcommand{\rmp}{Rev.\ Mod.\ Phys.}
\newcommand{\ssr}{Space Sci.\ Rev.}
\newcommand{\pasp}{PASP}
\newcommand{\nar}{New Astronomy Reviews}
\newcommand{\araa}{Annual Review of Astronomy and Astrophysics}

\end{document}